\documentclass[12pt,a4paper]{article}
\usepackage[utf8x]{inputenx}
\usepackage{ucs}
\usepackage[english]{babel}
\usepackage{amsmath}
\usepackage{amssymb}
\usepackage{amsfonts}
\usepackage{amsthm}
\usepackage{amscd}
\usepackage{amsbsy}
\usepackage{mathabx}
\usepackage{pgfplots}
\usepackage{fancyhdr}
\usepackage{xspace}
\usepackage{pdfpages}
\usepackage{mathrsfs}
\usepackage{pdfpages}
\usepackage{graphicx}
\usepackage{subfig}
\usepackage{xcolor}
\usepackage{ulem}
\usepackage{authblk}
\usepackage{float}
\numberwithin{equation}{section}

\theoremstyle{definition}
\newtheorem{remark}{Remark}[section]

\usepackage[bookmarks=true]{hyperref}
\usepackage{bookmark}

\newcommand*{\diff}{\mathop{}\!\mathrm{d}}
\title{A sub-Riemannian model of the functional architecture of M1 for arm movement direction}
\date{}
\author[1,2]{C. Mazzetti 
\thanks{caterina.mazzetti2@unibo.it}}
\author[2]{A. Sarti\thanks{alessandro.sarti@ehess.fr}}
\author[1,2]{G. Citti\thanks{giovanna.citti@unibo.it}}
\affil[1]{\small Department of Mathematics, University of Bologna}
\affil[2]{Centre d'Analyse et de Math\'{e}matique Sociales, Sorbonne Universit\'{e}}
\begin{document}
\maketitle
%
%
\begin{abstract}
In this paper we propose a neurogeometrical model of the behaviour of cells of the arm area of the primary motor cortex (M1). We mathematically express the hypercolumnar organization of M1 discovered by Georgopoulos, as a fiber bundle, as in classical sub-riemannian models of the visual cortex (Hoffmann, Petitot, Citti-Sarti).
On this structure, we consider the selective tuning of M1 neurons of kinematic variables of positions and directions of movement. We then extend this model to encode the notion of fragments of movements introduced by Hatsopoulos.  In our approach fragments are modelled as integral curves of vector fields in a suitable sub-Riemannian space. These fragments are in good agreements with movement decomposition from neural activity data. Here, we recover these patterns through a spectral clustering algorithm in the subriemannian structure we introduced, and compare our results with the neurophysiological ones of Kadmon-Harpaz et al.\\

\end{abstract}
\section{Introduction}
A fundamental problem regarding the study of motor cortex deals with the information conveyed by the discharge pattern of motor cortical cells.
This is a quite difficult topic  as the input of primary motor area comes from higher brain cortical regions, whereas the output is movement.

\begin{figure}[H]
\centering
\subfloat[]{\includegraphics[scale= 0.3]{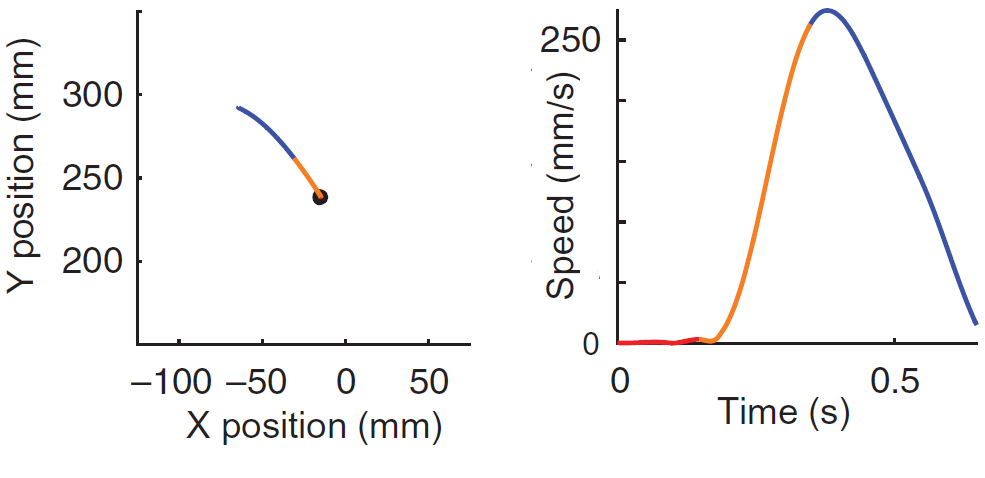}}\quad\quad
\subfloat[]{\includegraphics[scale= 0.3]{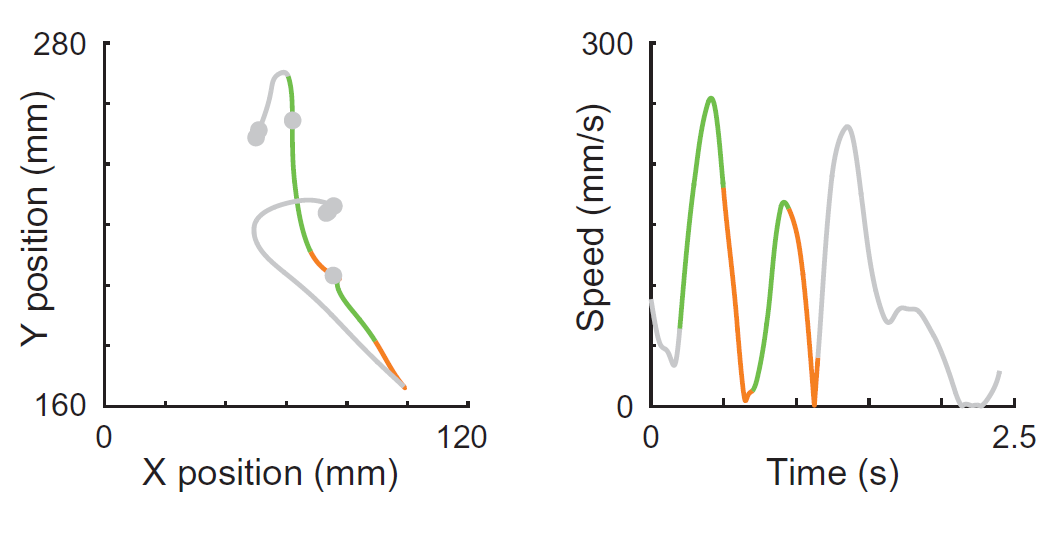}}
\caption{\small Examples of a center-out task and of a random target pursuit task, with position and speed profile colored according to the identified neural states. Black dot represents the starting position. From \cite{kadmon2019movement}.}
\label{fig_ret}
\end{figure}

Starting from 1978, a pioneering work for the study of primary motor cortex (M1) was developed by A. Georgopoulos, whose experiments allow 
to recognize many important features  of the arm area functional architecture. In particular, he discovered that cells of this area are 
sensible to the position and direction of the hand movement (\cite{georgopoulos1984static}, \cite{georgopoulos1982relations}), and are organized in a columnar structure, according to movement directions \cite{georgopoulos2015columnar}.
After the work of Georgopoulos, other experiments proved that activity of neurons in M1 correlates with a
broader variety of movement-related variables, including endpoint position, velocity, acceleration  (see \cite{schwartz2007useful} as a review). 
This phenomenon, known as ``cortical tuning", describes the selective responsiveness of M1 neurons to specific movement features. In other words, M1 neurons become active or ``tuned" based on specific movement characteristics.
Furthermore, the tuning for movement parameters is not static, but varies with time \cite{paninski2004spatiotemporal} and 
for this reason Hatsopoulos et al \cite{Encoding} proved that individual motor cortical cells rather encode short movement trajectories, called ``movement fragments" (see Fig. 2 (a)). 
Comparable findings were obtained by Kadmon-Harpaz et al \cite{kadmon2019movement} in 2019 who 
examined the temporal dynamics of neural populations in the primary motor cortex of macaque monkeys performing forelimb reaching movements. Using a
hidden Markov model, they found a structure of hidden states in the population activity of neurons in M1, which organizes  the behavioural output, in acceleration and deceleration trajectory segments with fixed directional selectivity (see Fig. \ref{fig_ret}). The data analysis was performed  at the neural level, and the authors posed the problem to recover the same decomposition by using only  kinematic variables.  

Aim of this paper is to answer to this problem, extending a result obtained  in \cite{Mazzetti}. We present a neurogeometrical model inspired by the functional architecture of the arm area of motor cortex referred to a set of cortical tuning parameters in response to point-to-point reaching movements. We modelled the time dependent selectivity of each neuron, through integral curves of a suitable sub-Riemannian vector fields in the space of kinematic variables. The same organization in elementary trajectories can be obtained with a kernel component analysis associated to the  sub-Riemannian distance. 

The main novelty with respect to \cite{Mazzetti} is that  we apply a second clustering in the space of elementary trajectories and we obtain, using only phenomenological  variables, the same neural PCAs provided by Karpaz-Harpaz et al \cite{kadmon2019movement}. 
In this second step the grouping is carried out in the space of curves (movement fragments), each one identified by its mean orientation and acceleration.  

The whole process provides detailed information on which kinematic variables are responsible for the neural process, but also gradually
leads to a shift from a space mainly described by kinematic points to a space of movement
trajectories.

\vspace{-\parskip}
\section{A sub-Riemannian model of M1 cells encoding movement direction}
We aim at realizing a unified neurogeometrical framework that contains both the geometrical findings of Georgopoulos regarding direction of movement and the time dipendent model of \cite{Encoding}. 
The space variables will be the  cortical features  of time, position, direction of movement, speed and acceleration, the constraint between them will be described via a  sub-Riemannian metric and time varying selective behaviour will be represented through integral curves of suitable vector fields. 

\subsection{A 2D kinematic tuning model of movement directions}\label{cortical_features_structure}

We first consider 
  that the basic functional properties of cellular activity in the arm area of M1 involve directional and positional tuning, 
 as described in  \cite{georgopoulos1982relations}, \cite{georgopoulos1984static}. Hence we introduce  
a variable $(x,y)$ which 
accounts for  hand's position in a two dimensional space, and a variable   
$\theta\in S^1$, which encodes hand's movement direction on the plane $(x,y)$. The relation between these variables 
is expressed by the relation $\frac{\diff y}{\diff x} = \tan(\theta)$, or equivalently by the vanishing of the 1-form 
\begin{equation}\label{citti_sarti}
\omega_{1} = -\sin\theta \diff x + \cos\theta \diff y= 0.
\end{equation}
It is worthwhile to note that this first constraint  is inspired by the models of visual cortex (see \cite{petitot1999vers}, \cite{citti2006cortical}).

We also consider that the cortex codes other movement-related variables, including velocity, acceleration  
(see \cite{schwartz2007useful}). Hence we introduce the time variable $t$, and   
the variables $v$ and $a$ which represent hand's speed and acceleration along the direction $\theta$. 

Recall that Georgopoulos \cite{georgopoulos2015columnar} also provided a physiological model of hypercolumnar organization for the cellular arrangement in M1. On the other hand, the hypercolumnar structure is also present in V1 where it has been modeled as a fiber bundle by Hoffmann \cite{hoffman1989visual} and Petitot and Tondut \cite{petitot1999vers}. This suggests to use a fiber bundle representation also as a model of M1. In our model, the triple $\left(t, x, y\right)\in\mathbb{R}^3$ describes the position of the hand at time $t$, and it is assumed to belong to the base space of a fiber bundle structure,  whereas the variables 
$\left(\theta, v, a\right)\in S^1 \times \mathbb{R}^{2}$ form the selected features on the fiber over the point $\left(t, x, y\right)$ (see \cite{Ledonne} for the definition of fiber bundle). We therefore consider the 6D features set 
\begin{equation}\label{6D}
\mathcal{M}= \mathbb{R}^{3}_{\left(t,x,y\right)} \times S^1_{\theta} \times \mathbb{R}^{2}_{\left(v,a\right)},
\end{equation}

Let us specify that in V1, the fiber bundle is compatible with the hypercolumnar organization and its correspondence with the retinal plane. Similarly, in M1, the fiber bundle captures the topographic organization resulting from competing mappings related to somatotopy, hand location, and movement organization \cite{graziano2007mapping}. Although there are differences in receptive profiles (simple cells in V1) versus ``actuator profiles" (cells in M1), both regions evaluate the alignment between preferred features and external input variables.

The choice of the space variables \eqref{6D} with their differential constraints induce the vanishing of the following 1-forms
\begin{equation}\label{tre_1forme}
\begin{aligned}
\omega_{2} = \cos\theta \diff x + \sin\theta \diff y - &v\diff t= 0,\quad
\omega_{3} &= \diff v -a\diff t= 0.
\end{aligned}
\end{equation}
The one-form $\omega_{2}$ encodes the direction of velocity over time: the unitary vector $\left(\cos\theta, \sin\theta\right)$ is the vector in the direction of velocity, and its product with $\left(\dot{x}, \dot{y}\right)$ yields the speed. We call horizontal distribution $D^{\mathcal{M}}$ 
the kernel of all three forms, which is the set of vector fields orthogonal to $\omega_i$, $i=1, ...3$. It turns out to be spanned by the vector fields
\begin{align}\label{campi_2D}
X_{1}= v\cos\theta\frac{\partial}{\partial{x}} + v\sin\theta\frac{\partial}{\partial{y}}+ a\frac{\partial}{\partial{v}}+ \frac{\partial}{\partial{t}},\quad
X_{2}= \frac{\partial}{\partial{\theta}},\quad
X_{3}= \frac{\partial}{\partial{a}}.
\end{align}
We call horizontal curve an integral curve of these vector fields: 
\begin{equation}\label{sistema_curve_integrali_polinomi}
\begin{cases}
\dot{\gamma}\left(t\right)= X_{1}\left(\gamma\left(t\right)\right)+ \dot{\theta}\left(t\right) X_{2}\left(\gamma\left(t\right)\right)+ \dot{a}\left(t\right)X_{3}\left(\gamma\left(t\right)\right)\\
\gamma\left(0\right)=\eta_0\in\mathcal{M}.
\end{cases}
\end{equation}
The functions $t\mapsto \dot{\theta}\left(t\right)$ and $t\mapsto \dot{a}\left(t\right)$ represent, respectively, the rate of change of the selective tuning to movement direction and acceleration variables. We identified each movement fragment detected in \cite{Encoding} as local integral curves of the space  $\mathcal{M}$ (Fig. 2 (a) and (b)). The 
whole space of cortical neurons is no more modelled as a set of points, but a set of trajectories, solution
of (\ref{sistema_curve_integrali_polinomi}) (Fig. 2 (c)).

\begin{figure}[H]
\centering
\begin{minipage}[b]{0.22\linewidth}
\centering
\subfloat[]{\includegraphics[scale=0.35]{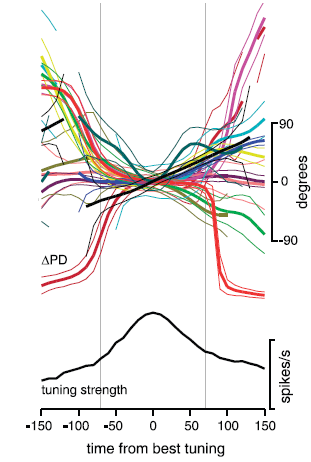}}
\end{minipage}\qquad
\begin{minipage}[b]{0.30\linewidth}
\centering
\subfloat[]{\includegraphics[scale=0.07]{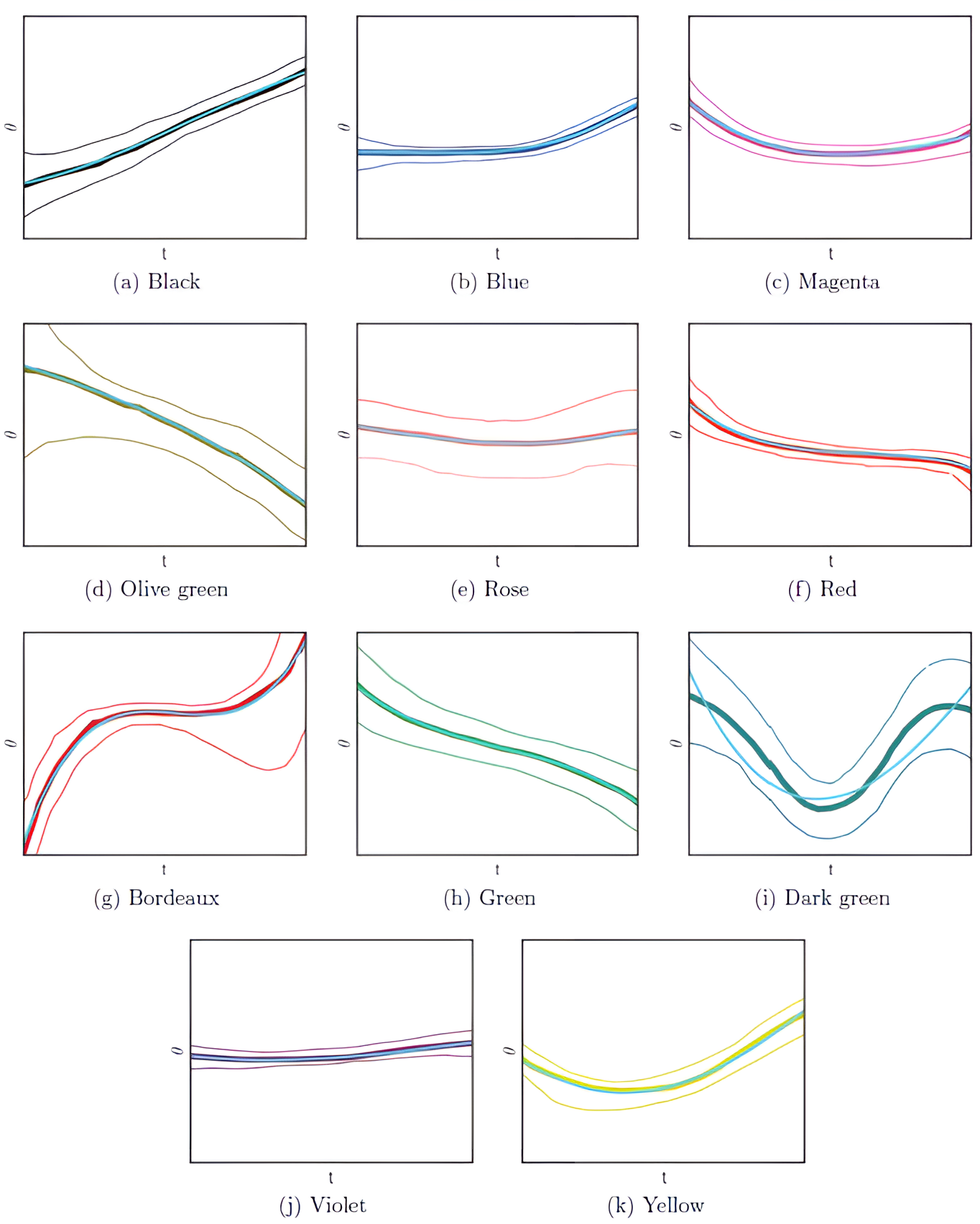}}
\end{minipage}
\begin{minipage}[b]{0.35\linewidth}
\centering
\subfloat[]{\includegraphics[scale=0.42]{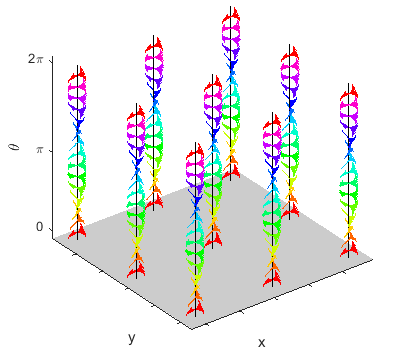}}
\end{minipage}
\caption{\small (a) Temporal evolution of the selective responsiveness to movement direction (direction tuning) in twelve neurons of M1. Below is shown the mean strength of direction tuning, where time 0 is assumed to be the time of strongest tuning. From \cite{Encoding}. 
(b) Best fit with an integral curve of the space (in blue) of each measured time-dependent direction tuning curve, represented with the same color as in \cite{Encoding}. (c) The cortical space is described as a space of integral curves of system \eqref{sistema_curve_integrali_polinomi}.}
\label{im_kinematic_bundle}
\end{figure}

\section{Spatio-temporal grouping model for M1}
We will now recover the coherent behaviors of neural activity obtained in \cite{kadmon2019movement} in terms of kinematic parameters. The authors noted that the desired neural decomposition 
 can not be obtained by none of the  distances previously proposed in literature. 
Here we show that a distance that takes into account the differential relations between the variables can provide the correct decomposition. The algorithm we will use to provide the classification is a variant of $k$-means that considers this distance, which proves that the set of kinematic variables considered is sufficient to recover the cortical decomposition.
\subsection{Homogeneous distance on $\mathcal{M}$}
Let us now introduce a natural distance associated to the  vector fields \eqref{campi_2D}, with the metric which makes them orthonormal. In this way we 
endow $\mathcal{M}$ with a sub-Riemannian structure.
Note that we only have chosen 3 vector fields at every point in a 6D space, and we will be able to obtain a basis of the space considering their commutators. We will also assign a degree, which is the number of  commutators we need to obtain a vector field.    Precisely 
$$X_1 ,\; X_2 ,\; X_3 \in D^{\mathcal{M}} \text{ so that } \text{deg}( X_1) =  \text{deg}( X_2) = \text{deg}( X_3)=1.$$ 
$$X_4 =[X_1, X_2],\; X_5 =[X_3, X_1], \text{ so that } \text{deg}(X_4) = \text{deg}(X_5) = 2$$ 
$$X_6 =[X_5, X_1] = [[X_3, X_1], X_1], \text{ so that } \text{deg}(X_4) = 3.$$
In particular the vector fields satisfy the H\"ormander condition, and a distance $d^{\mathcal{M}}$ 
(called Carnot-Carathéodory distance) can be defined as follows 
in the whole cortical feature space $\mathcal{M}$. For all  $\eta_0, \eta_1\in\mathcal{M}$ 
\begin{equation}\label{distanza_in_M}
d^\mathcal{M}\left(\eta_0, \eta_1\right)= \inf\left\{l\left(\gamma\right): \gamma\; \text{is a horizontal curve connecting}\: \eta_0\; \text{and}\; \eta_1\right\},
\end{equation}
where the notion of horizontal curve has been introduced in \eqref{sistema_curve_integrali_polinomi}. 
The path that realizes \eqref{distanza_in_M}
is called geodesic. 
Geodesics in this space are related to a model of arm-reaching movements proposed by Flash and Hogan \cite{FH} (see \cite{mazzetti2023model} for further details).\\
It is possible to provide a local estimate of the distance $d^{\mathcal{M}}$ using an approximation result due to Nagel Stein Wainger \cite{nagel1985balls}. 
\begin{remark}\label{def_distanza_hom}
We fix a point $\eta_0 =\left(x_{0},y_{0},\theta_{0},v_{0},a_{0},t_{0}\right)$ and call canonical coordinates of any other point 
$\eta_1= \left(x_{1},y_{1},\theta_{1},v_{1},a_{1},t_{1}\right)$ the constants $e_i$  which solve the system
\begin{align}\label{sistema_coord_esponenziali}
\begin{cases}
\dot{\gamma}\left(s\right)= e_{1}X_{1} + e_{2} X_{2}+ e_{3}X_{3}+ e_{4}X_{4}+ e_{5}X_{5}+ e_{6}X_{6}\\
\gamma\left(0\right)= \eta_0\quad 
\gamma\left(1\right)=\eta_1.\\
\end{cases}
\end{align}
Given a family of constant positive coefficients $\lbrace c_i\rbrace_{i=1}^{6}$, according to the work of Nagel et al. \cite{nagel1985balls}, the homogeneous distance between two points $\eta_0$, $\eta_1\in \mathcal{M}$ can be estimated as follows
\begin{equation}\label{distanza_exp_6D}
d^{\mathcal{M}} \left(\eta_0,\eta_1\right)\approx
\left(\sum_{i=1}^6 c_i\left|e_{i}\right|^{6/\text{deg}(X_i)} \right)^{\frac{1}{6}}.
\end{equation}
\end{remark}
In our experiments, we will use this estimate of the distance. 
\subsection{Model of movement decomposition}
\subsubsection{Clustering for identification of elementary trajectories \newline}\label{fragment_identif}
We define in the cortical feature space  $\mathcal{M}$, a connectivity kernel $\omega_{\mathcal{M}}$ expressed in term of distance \eqref{distanza_exp_6D}: 
\begin{equation}
\label{kernel_m_capitolo_grouping}
\omega_{\mathcal{M}}\left(\eta_0,\eta\right)= e^{- d^{\mathcal{M}}\left(\eta_0,\eta\right)^2}, \quad \eta_0, \eta\in \mathcal{M},
\end{equation}
Then, we discretize \eqref{kernel_m_capitolo_grouping} 
on  a set of reaching paths, and obtain a real symmetric affinity matrix $A$:
 
\begin{equation}\label{affinity_matrix_tesi}
A= \omega_{\mathcal{M}}\left(\left(x_i, y_i, \theta_i, v_i, a_i, t_i\right), \left(x_j, y_j, \theta_j, v_j, a_j, t_j\right)\right),
\end{equation}
which contains the connectivity information between all the
kinematic variables of the reaching trajectory. 

We propose to apply a spectral clustering 
analysis of this affinity matrix to a set of movement trajectories 
in order to obtain a decomposition in elementary trajectories to be compared with  
the one obtained in \cite{kadmon2019movement} (see Figure \ref{fig_ret} as a reference). 

\subsubsection{Clustering for classification of elementary trajectories\newline}\label{fragment_classif}
The next step involves grouping the elementary trajectories based on the properties described in Kadmon Harpaz et al \cite{kadmon2019movement}. 
Recall that elementary trajectories are regular curves with values 
in the space $\mathcal{M}$. Up to a change of e parameterization in the variable $s$, we can assume that the elementary trajectory space $\mathcal{F}\left(\mathcal{M}\right)$ is a subset of 
$C^1([0,1], \mathcal{M})$.

In order to perform a grouping of these elementary trajectories we note that  
movement direction $\theta$ and acceleration $a$ are almost constant on the elementary trajectories recovered in sections \ref{fragment_identif}, 
and that neurons are invariant with respect to time and position (e.g. \cite{Encoding}, \cite{kadmon2019movement}). 

Hence we associate to each elementary trajectory its mean orientation and acceleration. If $\gamma:[0,1]\to \mathcal{M}$ is an elementary trajectory, we denote 
$$\bar{\theta}(\gamma) = \int_0^1 \theta(s) ds \;\;  \text{and }\;\; \bar{a}(\gamma)= \int_0^1 a(s) ds.
$$


Subsequently, we perform a new clustering by adapting the sub-Riemannian distance previously defined. 
Precisely, if $\gamma_0,$$\gamma_1$ are elementary trajectories, the new distance will be defined

$$d^{\mathcal{F}}(\gamma_0, \gamma_1) = d^{\mathcal{M}}\Big(\big(0,0,\bar \theta(\gamma_0),0, \bar a(\gamma_0), 0\big), \big(0,0,\bar \theta(\gamma_1),0, \bar a(\gamma_1),0\big)\Big).$$

The  kernel over the elementary trajectories space is given by 
\begin{align}
\omega_{\mathcal{F}}\left(\gamma_0,\gamma_1\right)= e^{- d^{\mathcal{F}}\left(\gamma_0,\gamma_1\right)^2}, \quad \gamma_0, \gamma_1\in \mathcal{F}\left(\mathcal{M}\right).
\end{align}
Affinity matrix turns out to be defined by consequence, and we can apply the previous clustering analysis using the distance between these pairs. 
Note that even if the elementary trajectories are curves, the clustering algorithm is performed on the averages, so that it takes place in a finite dimensional space. 
In this way we obtain a classification of elementary trajectories into classes, called fragments, to be  compared with the ones in  Kadmon-Harpaz et al \cite{kadmon2019movement}.

\subsection{Results}

In the following we will show two test cases. In both cases, as in the paper \cite{kadmon2019movement}, the motion trajectory  is visualized by two graphs, one on the $\left(x,y\right)$ plane, the reaching path, and one on the $\left(t,v\right)$ plane corresponding to the velocity profile.
In Test 1, the clustering method described in \cite{Mazzetti} is sufficient. However, in Test 2, in order to achieve the classification results obtained by Kadmon-Harpaz et al. \cite{kadmon2019movement}, it is necessary to utilize the newly adopted algorithm outlined in section \ref{fragment_classif}.
\subsubsection*{Test 1: Simulation of a center-out task\newline}
As a first example, we will analyze a trajectory of movement performing a center-out task. 
\begin{figure}[H]
\centering
\subfloat[]{\includegraphics[scale= 0.2]{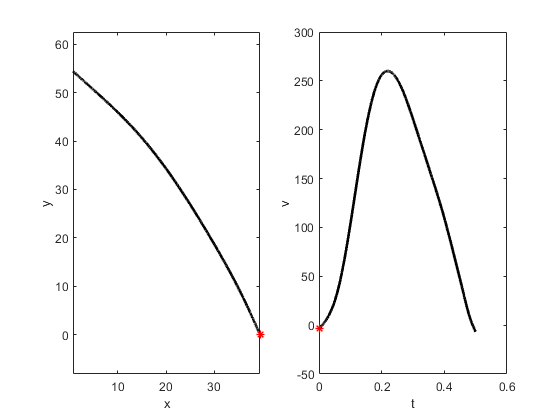}}
\subfloat[]{\includegraphics[scale=0.2]{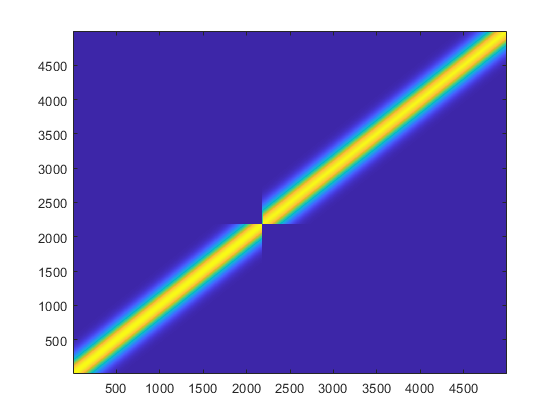}}
\subfloat[]{\includegraphics[scale= 0.2]{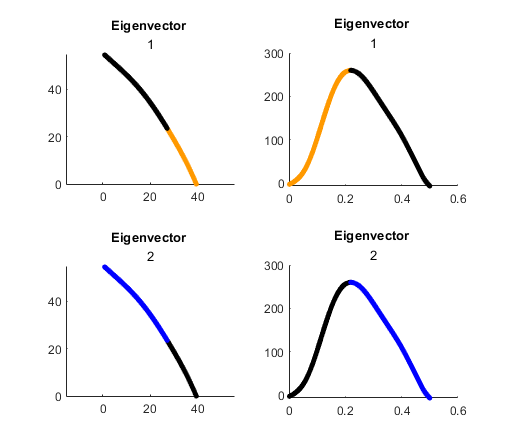}}
\subfloat[]{\includegraphics[scale= 0.2]{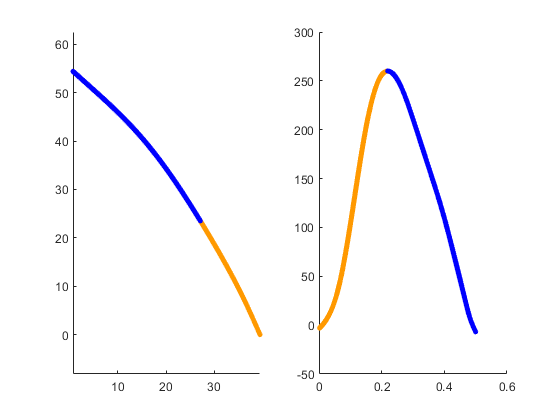}}
\caption{\small (a) Reaching path and speed profile of a center-out task over the $\left(x,y\right)$ plane. and $\left(t,v\right)$ plane. The red dot represents the movement starting position. (b) The Affinity matrix.  (c-d) Projections of the eigenvectors.}
\label{approx_fig_ret}
\end{figure}

In this very simple case, movement direction is almost constant with only one target point to be reached and just one maximum point is present on the speed profile. The affinity matrix which is clearly divided in two blocks (Fig. \ref{approx_fig_ret} (b)) identifies the eigenspaces associated to the two major eigenvalues. The projection of the  eigenvectors over the reaching trajectory ( Fig. \ref{approx_fig_ret} (c) and (d))), corresponds precisely to the acceleration and deceleration phases of the movement task coherently with the neural states found in \cite{kadmon2019movement} (see also Fig. \ref{fig_ret} (a)). In this simple case we do not need to apply the second clustering of our algorithm. 


\subsubsection*{Test 2: simulation of a random target pursuit task\newline}
\begin{figure}[H]
\centering
\subfloat[]{\includegraphics[scale=0.2]{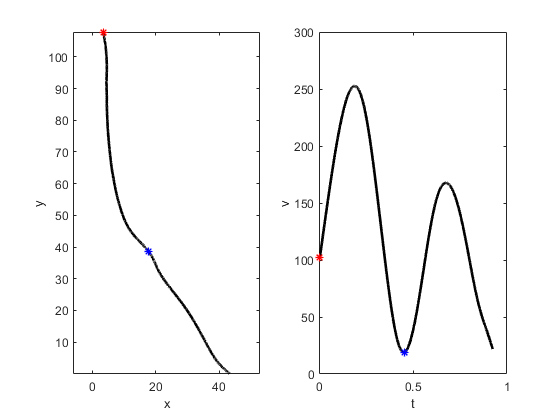}}
\subfloat[]{\includegraphics[scale=0.2]{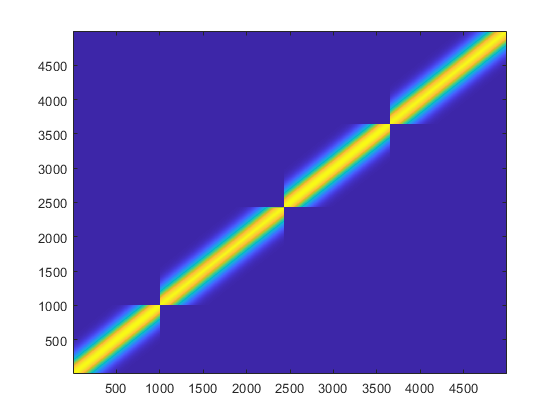}}
\subfloat[]{\includegraphics[scale=0.2]{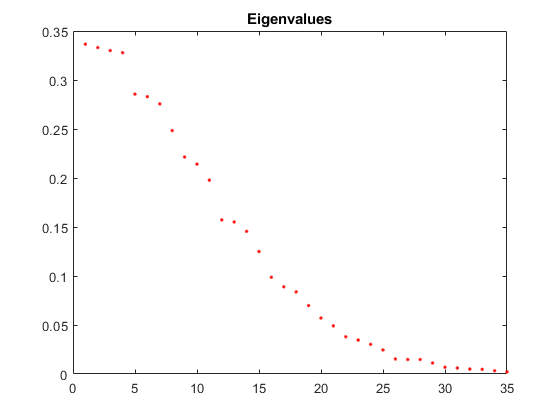}}
\subfloat[]{\includegraphics[scale=0.2]{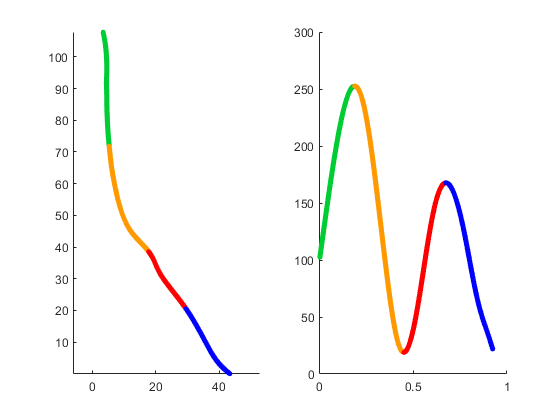}}
\caption{\small Reaching path and speed profile of a random target pursuit task: approximation of Fig. \ref{fig_ret} (b). (b-c) The Affinity matrix and the eigenvalues plot. (d) Projections of the eigenvectors over the reaching trajectory.}
\label{approx_fig_4}
\end{figure}

In this test, we apply our spectral algorithm on an approximate trajectory of Fig. \ref{fig_ret} (b). The analyzed motion is represented in Fig. \ref{approx_fig_4} (a). 
In \cite{kadmon2019movement}, the experiment performed by the monkey consists of reaching several targets one after the other. Here, a trajectory is extrapolated that starts from a fixed point (red point in Fig. \ref{approx_fig_4} (a)), arrives at a second target (blue point in Fig. \ref{approx_fig_4} (a)) and comes to an end.
The affinity matrix is divided into four blocks (see Fig. \ref{approx_fig_4} (b)). As before we project the eigenvectors associated with the largest eigenvalues onto the motion trajectory (see Fig. \ref{approx_fig_4} (d)). 

\begin{figure}[H]
\centering
\subfloat[]{\includegraphics[scale= 0.25]{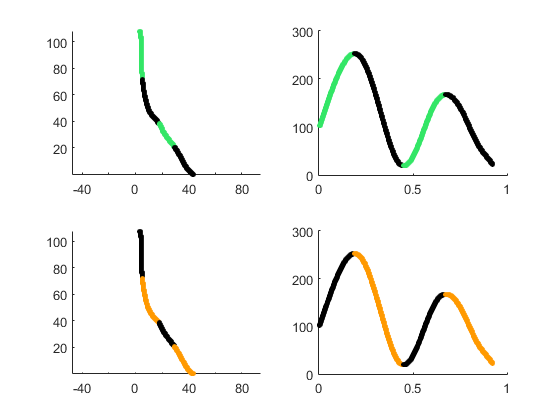}}\quad
\subfloat[]{\includegraphics[scale=0.25]{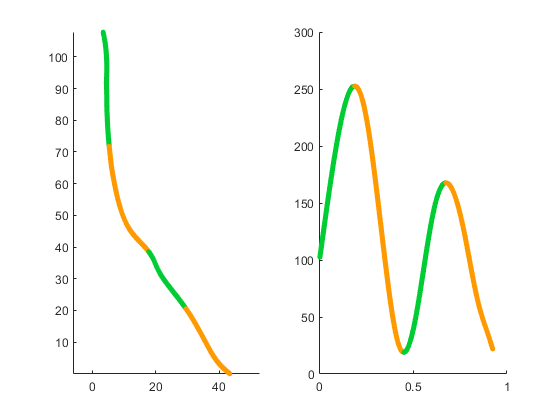}}
\caption{\small Resulting reaching trajectory segmentation according to spatio-temporal invariant clusters. Acceleration and deceleration phases are respectively identified. See Fig. \ref{fig_ret} (b) for a direct comparison.}
\label{global_dec_fig_4}
\end{figure}

After that we apply the second clustering in the space of sub-trajectories,  with respect to the $\theta$ and $a$ variables.
 The resulting clusters appropriately group acceleration and
deceleration phases, as well as phases with constant direction. The eigenvectors colored in green denote the acceleration phase, those colored in orange the deceleration phase. 
The resulting decomposition pattern displayed in Fig. \ref{global_dec_fig_4} is in agreement with the
experimental result of \cite{kadmon2019movement} shown in Fig. \ref{fig_ret} (b).

\section{Conclusions}
We presented a sub-Riemannian model of the arm area of motor cortex 
expressed in terms of the kinematic variables experimentally measured in 
\cite{georgopoulos1984static} and \cite{schwartz2007useful}. The metric of the space was directly deduced from the constraint between these variables, 
and was expressed in terms of suitable vector fields. We showed that their integral curves 
provide a good model of the time-dependent directional tuning of neurons in this area, experimentally found in \cite{Encoding}. 
We finally introduced a distance that allows to perform a kernel component analysis which is the phenomenological counterpart of the neural PCAs provided by Kadmon-Harpaz et al \cite{kadmon2019movement}. 
In particular, we emphasize that by working only on kinematic variables we recovered the same neural classification acquired by electrode array. 
This proves that the distance $d^{\mathcal{M}}$ is adequate, not only because of the properties of the kinematic space, but also because of the classification in sub-trajectory fragments given by the clustering algorithm, which has a neural foundation. In particular the kinematic parameters we identified are sufficient to completely explain the process observed in \cite{kadmon2019movement}.\\
\\
\small
\noindent\small\textbf{Acknowledgments:} \small GHAIA project, H2020 MSCA  n. 777622;  NGEU-MUR-NRRP, project MNESYS n. PE0000006. 

\bibliographystyle{splncs04}
\bibliography{mybibliography}

\small
\begin{thebibliography}{8}
\bibitem{churchland2007temporal}
M. Churchland and K. Shenoy.
\newblock Temporal complexity and heterogeneity of single-neuron activity in
  premotor and motor cortex.
\newblock {\em J. of neuroph.}, 97(6):4235--57, 2007.

\bibitem{citti2006cortical}
 G. Citti, A. Sarti.
 \newblock A cortical based model of perceptual completion in the
   roto-translation space.
 \newblock {\em J. of Math. Imag. Vis.}, 24(3):307--326,
  2006.


\bibitem{FH}
T. Flash and N. Hogan.
\newblock The coordination of arm movements: an experimentally confirmed
  mathematical model.
\newblock {\em J. of neuroscience}, 5(7):1688--1703, 1985.


\bibitem{georgopoulos1984static}
A.P. Georgopoulos, R. Caminiti, and J.F. Kalaska.
\newblock Static spatial effects in motor cortex and area 5: quantitative
  relations in a two-dimensional space.
\newblock {\em Exp. Brain Research}, 54(3):446--454, 1984.

\bibitem{georgopoulos2015columnar}
A.P. Georgopoulos.
\newblock Columnar organization of the motor cortex: direction of movement.
\newblock In {\em Rec. Adv. on the Mod. Org. of the Cortex}, 123--141. Springer, 2015.

\bibitem{georgopoulos1982relations}
A.P. Georgopoulos, J.F. Kalaska, R. Caminiti, and J.T. Massey.
\newblock On the relations between the direction of two-dimensional arm
  movements and cell discharge in primate motor cortex.
\newblock {\em J. of Neuroscience}, 2(11):1527--1537, 1982.

\bibitem{graziano2007mapping}
Michael~SA Graziano and Tyson~N Aflalo.
\newblock Mapping behavioral repertoire onto the cortex.
\newblock {\em Neuron}, 56(2):239--251, 2007.


\bibitem{Encoding}
N.G. Hatsopoulos, Q. Xu, and Y. Amit.
\newblock Encoding of movement fragments in the motor cortex.
\newblock {\em J. of Neuroscience}, 27(19):5105--5114, 2007.


\bibitem{hoffman1989visual}
W. C. Hoffmann, 
\newblock The visual cortex is a contact bundle.
\newblock {\em Applied Mathematics and Computation}, 32:132--167, 1989.

\bibitem{kadmon2019movement}
N. Kadmon Harpaz, D. Ungarish, N.G. Hatsopoulos, and T. Flash.
\newblock Movement decomposition in the primary motor cortex.
\newblock {\em Cerebral cortex}, 29(4):1619--1633, 2019.

\bibitem{kettner1988primate}
R.E. Kettner, A.B. Schwartz, and A.P. Georgopoulos.
\newblock Primate motor cortex and free arm movements to visual targets in
  three-dimensional space. iii. 
\newblock {\em J. of Neuroscience}, 8(8):2938--2947, 1988.

\bibitem{Ledonne}
E. Le~Donne.
\newblock Lecture notes on sub-riemannian geometry.
\newblock {\em preprint}, 2010.

\bibitem{Mazzetti}
C. Mazzetti, A. Sarti, G. Citti
\newblock Functional architecture of M1 cells encoding movement direction.
\newblock {\em arXiv preprint arXiv:2208.03352}, 2022.


\bibitem{mazzetti2023model}
C. Mazzetti, A. Sarti, G. Citti
\newblock A model of reaching via subriemannian geodesics in Engel-type group.
\newblock {\em arXiv preprint arXiv:2301.05765}, 2023.

\bibitem{nagel1985balls}
A. Nagel, E.M. Stein, and S. Wainger.
\newblock Balls and metrics defined by vector fields i: Basic properties.
\newblock {\em Acta Mathematica}, 155:103--147, 1985.

\bibitem{paninski2004spatiotemporal}
L. Paninski, M.R. Fellows, N.G. Hatsopoulos, and J.P. Donoghue.
\newblock Spatiotemporal tuning of motor cortical neurons for hand position and
  velocity.
\newblock {\em J. of neuroph.}, 91(1):515--532, 2004.

\bibitem{petitot1999vers}
 J. Petitot and Y. Tondut.
 \newblock Vers une neurog{\'e}om{\'e}trie. fibrations corticales, structures de
   contact et contours subjectifs modaux.
 \newblock {\em Math{\'e}matiques et sciences humaines}, 145:5--101, 1999.

\bibitem{schwartz2007useful}
A.B. Schwartz.
\newblock Useful signals from motor cortex.
\newblock {\em J. of phys.}, 579(3):581--601, 2007.

\end{thebibliography}

\end{document}